\documentclass[aps,rpl,preprint,groupedaddress]{revtex4}
\begin{document}

\title{Time correlated quantum amplitude damping channel}
\author{Ye Yeo and Andrew Skeen}

\affiliation{Centre for Mathematical Sciences, Wilberforce Road, Cambridge CB3 0WB, United Kingdom}

\begin{abstract}
We analyze the problem of sending classical information through qubit channels where successive uses of the channel are correlated.  This work extends the analysis of C. Macchiavello and G. M. Palma to the case of a non-Pauli channel - the amplitude damping channel.  Using the channel description outlined in S. Daffer, {\it et al.}, we derive the correlated amplitude damping channel.  We obtain a similar result to C. Macchiavello and G. M. Palma, that is, that under certain conditions on the degree of channel memory, the use of entangled input signals may enhance the information transmission compared to the use of product input signals.
\end{abstract}

\maketitle

In real physical quantum transmission channels, it is common to have correlated noise acting on consecutive uses.  This is in contrast to the class of memoryless channels, in which uncorrelated (independent) noise acts on each use.  The problem of the classical capacity of quantum channels with time correlated noise was first considered by C. Macchiavello and G. M. Palma \cite{Mac}.  They analyzed the specific case of sending qubits (quantum states belonging to 2-dimensional Hilbert spaces, each spanned by orthonormal vectors $\{|0\rangle,\ |1\rangle\}$) with two consecutive uses of a quantum depolarizing channel with partial memory.  The action of such a channel on an input state described by the density operator $\pi$, is defined by the completely positive, trace-preserving map to an output state, another density operator $\rho$ \cite{Kraus, Mac}:
$$
\pi \longrightarrow \rho = \Phi(\pi) = (1 - \mu)\sum^3_{i, j = 0}A^u_{ij}\pi A^{u\dagger}_{ij} + \mu\sum^3_{k = 0}A^c_{kk}\pi A^{c\dagger}_{kk},
$$
where $0 \leq \mu \leq 1$.  With probability $(1 - \mu)$, the noise is uncorrelated and completely sepcified by the Kraus operators
$$
A^u_{00} = p_0I \otimes I,\
A^u_{01} = \sqrt{p_0p_1}I \otimes \sigma_x,\
A^u_{02} = \sqrt{p_0p_2}I \otimes \sigma_y,\
A^u_{03} = \sqrt{p_0p_3}I \otimes \sigma_z,
$$
$$
A^u_{10} = \sqrt{p_0p_1}\sigma_x \otimes I,\
A^u_{11} = p_1\sigma_x \otimes \sigma_x,\
A^u_{12} = \sqrt{p_1p_2}\sigma_x \otimes \sigma_y,\
A^u_{13} = \sqrt{p_1p_3}\sigma_x \otimes \sigma_z,
$$
$$
A^u_{20} = \sqrt{p_0p_2}\sigma_y \otimes I,\
A^u_{21} = \sqrt{p_1p_2}\sigma_y \otimes \sigma_x,\
A^u_{22} = p_2\sigma_y \otimes \sigma_y,\
A^u_{23} = \sqrt{p_2p_3}\sigma_y \otimes \sigma_z,
$$
$$
A^u_{30} = \sqrt{p_0p_3}\sigma_z \otimes I,\
A^u_{31} = \sqrt{p_1p_3}\sigma_z \otimes \sigma_x,\
A^u_{32} = \sqrt{p_2p_3}\sigma_z \otimes \sigma_y,\
A^u_{33} = p_3\sigma_z \otimes \sigma_z,
$$
while with probability $\mu$, the noise is correlated and specified by
$$
A^c_{00} = \sqrt{p_0}I \otimes I,\
A^c_{11} = \sqrt{p_1}\sigma_x \otimes \sigma_x,\
A^c_{22} = \sqrt{p_2}\sigma_y \otimes \sigma_y,\
A^c_{33} = \sqrt{p_3}\sigma_z \otimes \sigma_z.
$$
Here, $0 \leq p \leq 1$, $p_0 = (1 - p)$, $p_1 = p_2 = p_3 = \frac{1}{3}p$, and
$$
I = \left(\begin{array}{cc}
1 & 0 \\ 0 & 1
\end{array}\right),\
\sigma_x = \left(\begin{array}{cc}
0 & 1 \\ 1 & 0
\end{array}\right),\
\sigma_y = \left(\begin{array}{cc}
0 & -i \\ i & 0
\end{array}\right),\
\sigma_z = \left(\begin{array}{cc}
1 & 0 \\ 0 & -1
\end{array}\right)
$$
are the identity and Pauli matrices respectively.  The key object in their analysis is the mutual information for two uses of the channel,
\begin{equation}
I_2({\cal E}) = S(\rho) - \sum_iq_iS(\rho_i)
\end{equation}
where ${\cal E} = \{q_i,\ \pi_i\}$, with $q_i \geq 0$, $\sum_iq_i = 1$, is the input ensemble of states $\pi_i$, transmitted with {\it a priori} probabilities $q_i$, of two - generally entangled - qubits.  In Eq.(1),
$$
S(\sigma) = -{\rm tr}(\sigma\log_2\sigma)
$$
is the von Neumann entropy, $\rho = \sum_iq_i\rho_i$, and $\rho_i = \Phi(\pi_i)$.  For the following choice of equally-weighted ensemble of orthonormal input states,
$$
\pi = \frac{1}{4}(|\pi_1\rangle\langle\pi_1| + |\pi_2\rangle\langle\pi_2| + |\pi_3\rangle\langle\pi_3| + |\pi_4\rangle\langle\pi_4|),
$$
$$
|\pi_1\rangle = \cos\theta|00\rangle + \sin\theta|11\rangle,
$$
$$
|\pi_2\rangle = \sin\theta|00\rangle - \cos\theta|11\rangle,
$$
$$
|\pi_3\rangle = \cos\theta|01\rangle + \sin\theta|10\rangle,
$$
\begin{equation}
|\pi_4\rangle = \sin\theta|01\rangle - \cos\theta|10\rangle,
\end{equation}
Eq.(1) reduces to
$$
I_2 = 2 + \sum^4_{i = 1}e_i\log_2e_i
$$
with
$$
\eta = 1 - \frac{4}{3}p,
$$
$$
e_1 = e_2 = \frac{1}{4}(1 - \eta^2)(1 - \mu),
$$
$$
e_{3, 4} = \frac{1}{4}\left[(1 + \mu) + \eta^2(1 - \mu) \pm 2\sqrt{\eta^2\cos^22\theta + [\mu + \eta^2(1 - \mu)]^2\sin^22\theta}\right].
$$
They showed that there exists a threshold value
$$
\mu_t = \frac{\eta}{1 + \eta}
$$
for $0 < \eta < 1$, such that when $\mu > \mu_t$, $I_2$ is maximal for $\theta = \frac{\pi}{4}$ (i.e., the maximally entangled Bell states), while when $\mu < \mu_t$, $I_2$ is maximal for $\theta = 0$ (i.e., the completely unentangled product states).  Furthermore, at $\mu = \mu_t$, any set of states of the form Eq.(2) leads to the same $I_2$.

We note that, in the same manner, a quantum dephasing channel with uncorrelated noise can be defined as one specified by the following Kraus operators
$$
D^u_{00} = (1 - p)I \otimes I,\ 
D^u_{01} = \sqrt{p(1 - p)}I \otimes \sigma_z
$$
\begin{equation}
D^u_{10} = \sqrt{p(1 - p)}\sigma_z \otimes I,\
D^u_{11} = p\sigma_z \otimes \sigma_z
\end{equation}
and one with correlated noise by
\begin{equation}
D^c_{00} = \sqrt{1 - p}I \otimes I,\
D^c_{11} = \sqrt{p}\sigma_z \otimes \sigma_z.
\end{equation}
The same prescription can be applied to a quantum amplitude damping channel with uncorrelated noise as one specified by
\begin{equation}
E^u_{00} = E_0 \otimes E_0,\ E^u_{01} = E_0 \otimes E_1,\
E^u_{10} = E_1 \otimes E_0,\ E^u_{11} = E_1 \otimes E_1,
\end{equation}
where, with $0 \leq \chi \leq \frac{\pi}{2}$,
\begin{equation}
E_0 = \left(\begin{array}{cc}
\cos\chi & 0 \\ 0 & 1
\end{array}\right),\
E_1 = \left(\begin{array}{cc}
0 & 0 \\ \sin\chi & 0
\end{array}\right)
\end{equation}
are the Kraus operators for an amplitude damping channel.  Here, $|0\rangle$ and $|1\rangle$ denote the excited and ground states respectively.  However, it is not {\it a priori} clear how the Kraus operators for a quantum amplitude damping channel with correlated noise could be constructed in a similar manner, if it is at all possible.

Recently, S. Daffer, {\it et al.} \cite{Daffer} used a special basis of left, $\{L_i\}$, and right, $\{R_i\}$, damping eigenoperators for a Lindblad superoperator,
$$
\Phi(\pi) = \exp(t{\cal L})\pi,
$$
where $t$ is time, to calculate explicitly the image of a completely positive, trace-preserving map for a wide class of Markov quantum channels:
\begin{equation}
\pi \longrightarrow \rho = \Phi(\pi) 
= \sum_i{\rm tr}(L_i\pi)\exp(\lambda_it)R_i.
\end{equation}
For a finite $N$-dimensional Hilbert space,
$$
{\cal L}\pi = -\frac{1}{2}\sum^{N^2 - 1}_{i, j = 1}c_{ij}(F^{\dagger}_jF_i\pi + \pi F^{\dagger}_jF_i - 2F_i\pi F^{\dagger}_j),
$$
with the system operators $F_i$ satisfying
$$
{\rm tr}(F_i) = 0,\ {\rm tr}(F^{\dagger}_iF_j) = \delta_{ij}.
$$
The complex $c_{ij}$ form a positive matrix.  The right eigenoperators $R_i$ satisfy the eigenvalue equation
\begin{equation}
{\cal L}R_i = \lambda_iR_i,
\end{equation}
and the following duality relation
\begin{equation}
{\rm tr}(L_iR_j) = \delta_{ij}
\end{equation}
with the left eigenoperators $L_i$.  The amplitude damping and dephasing channels are examples of quantum Markov channels.  The Lindblad equation \cite{Daffer}
\begin{equation}
{\cal L}\pi = -\frac{1}{2}\alpha(\sigma^{\dagger}\sigma\pi + \pi\sigma^{\dagger}\sigma - 2\sigma\pi\sigma^{\dagger}),
\end{equation}
where $\alpha$ is a parameter analogous to the Einstein coefficient of spontaneous emission, and
$$
\sigma^{\dagger} \equiv \frac{1}{2}(\sigma_x + i\sigma_y),\
\sigma \equiv \frac{1}{2}(\sigma_x - i\sigma_y)
$$
are the creation and annihilation operators respectively, yields the amplitude damping channel, Eq.(6).  And, the dephasing channel can be derived from \cite{Daffer}
\begin{equation}
{\cal L}\pi = -\frac{1}{2}\Gamma(\pi - \sigma_z\pi\sigma_z),
\end{equation}
where $\Gamma$ is another parameter.  In this paper, we derive Eq.(4) from the Lindblad equation Eq.(12), that is, it gives the quantum dephasing channel with correlated noise.  In a similar fashion, we solve Eq.(19) and interpret the resulting completely positive, trace-preserving map as one which describes a quantum amplitude damping channel with correlated noise.  We then analyze, as in Ref.\cite{Mac}, the action of a quantum amplitude damping channel with partial memory on Eq.(2).  Our results are in agreement with those of Ref.\cite{Mac}.  That is, the transmission of classical information can be enhanced by employing maximally entangled states as carriers of information rather than product states.

We begin by solving the following Lindblad equation
\begin{equation}
{\cal L}\pi = -\frac{1}{2}\Gamma[\pi - (\sigma_z \otimes \sigma_z)\pi(\sigma_z \otimes \sigma_z)].
\end{equation}
Eq.(12) is an obvious extension of Eq.(11) with $\sigma_z$ replaced by $(\sigma_z \otimes \sigma_z)$.  The rationale is so that the phase-flip actions of the channel would then be correlated.  The method of solution involves first determining the right eigenoperators $R_i$, which solves Eq.(8):
\begin{equation}
R_{00} = \frac{1}{\sqrt{2}}\left(\begin{array}{cccc}
1 & 0 & 0 & 0 \\
0 & 0 & 0 & 0 \\
0 & 0 & 0 & 0 \\
0 & 0 & 0 & 1
\end{array}\right),\ \lambda_{00} = 0;\
R_{33} = \frac{1}{\sqrt{2}}\left(\begin{array}{cccc}
1 & 0 & 0 & 0 \\
0 & 0 & 0 & 0 \\
0 & 0 & 0 & 0 \\
0 & 0 & 0 & -1
\end{array}\right),\ \lambda_{33} = 0;
\end{equation}
\begin{equation}
R_{01}^{\pm} = \frac{1}{\sqrt{2}}\left(\begin{array}{cccc}
0 & \pm 1 & 0 & 0 \\
1 & 0 & 0 & 0 \\
0 & 0 & 0 & 0 \\
0 & 0 & 0 & 0
\end{array}\right),\ \lambda_{01}^{\pm} = -\Gamma;\
R_{02}^{\pm} = \frac{1}{\sqrt{2}}\left(\begin{array}{cccc}
0 & 0 & \pm 1 & 0 \\
0 & 0 & 0 & 0 \\
1 & 0 & 0 & 0 \\
0 & 0 & 0 & 0
\end{array}\right),\ \lambda_{02}^{\pm} = -\Gamma;
\end{equation}
\begin{equation}
R_{03}^{\pm} = \frac{1}{\sqrt{2}}\left(\begin{array}{cccc}
0 & 0 & 0 & \pm 1 \\
0 & 0 & 0 & 0 \\
0 & 0 & 0 & 0 \\
1 & 0 & 0 & 0
\end{array}\right),\ \lambda_{03}^{\pm} = 0;\
R_{11} = \left(\begin{array}{cccc}
0 & 0 & 0 & 0 \\
0 & 1 & 0 & 0 \\
0 & 0 & 0 & 0 \\
0 & 0 & 0 & 0
\end{array}\right),\ \lambda_{11} = 0;
\end{equation}
\begin{equation}
R_{12}^{\pm} = \frac{1}{\sqrt{2}}\left(\begin{array}{cccc}
0 & 0 & 0 & 0 \\
0 & 0 & \pm 1 & 0 \\
0 & 1 & 0 & 0 \\
0 & 0 & 0 & 0
\end{array}\right),\ \lambda_{12}^{\pm} = 0;\
R_{13}^{\pm} = \frac{1}{\sqrt{2}}\left(\begin{array}{cccc}
0 & 0 & 0 & 0 \\
0 & 0 & 0 & \pm 1 \\
0 & 0 & 0 & 0 \\
0 & 1 & 0 & 0
\end{array}\right),\ \lambda_{13}^{\pm} = -\Gamma;
\end{equation}
\begin{equation}
R_{22} = \left(\begin{array}{cccc}
0 & 0 & 0 & 0 \\
0 & 0 & 0 & 0 \\
0 & 0 & 1 & 0 \\
0 & 0 & 0 & 0
\end{array}\right),\ \lambda_{22} = 0;\
R_{23}^{\pm} = \frac{1}{\sqrt{2}}\left(\begin{array}{cccc}
0 & 0 & 0 & 0 \\
0 & 0 & 0 & 0 \\
0 & 0 & 0 & \pm 1\\
0 & 0 & 1 & 0
\end{array}\right),\ \lambda_{23}^{\pm} = -\Gamma.
\end{equation}
Notice that the index $i$ here is a doublet.  Next, the left eigenoperators are determined by imposing Eq.(9).  Finally, the image of the completely positive, trace-preserving map can be obtained via Eq.(7).  In this case, we have
$$
\pi \longrightarrow \rho = \sum_i{\rm tr}(L_i\pi)\exp(\lambda_it)R_i
= \sum^1_{j = 0}D^c_{jj}\pi D^{c\dagger}_{jj},
$$
where $D^c_{jj}$ are given by Eq.(4), with
\begin{equation}
p \equiv \frac{1}{2}[1 - \exp(-\Gamma t)].
\end{equation}
Therefore, Eq.(12) does indeed yield a dephasing channel with correlated noise.

Next, we solve the Lindblad equation
\begin{equation}
{\cal L}\pi = -\frac{1}{2}\alpha[(\sigma^{\dagger}\otimes\sigma^{\dagger})(\sigma\otimes\sigma)\pi + \pi(\sigma^{\dagger}\otimes\sigma^{\dagger})(\sigma\otimes\sigma) - 2(\sigma\otimes\sigma)\pi(\sigma^{\dagger}\otimes\sigma^{\dagger})].
\end{equation}
This follows from the same rationale behind the construction of Eq.(12).  By replacing $\sigma$ in Eq.(10) with $(\sigma \otimes \sigma)$, we expect the actions of the resulting channel to be correlated.  We call it the quantum amplitude damping channel with correlated noise.  The right eigenoperators $R_i$, which solves Eq.(8) are
\begin{equation}
R_{00} = \frac{1}{\sqrt{2}}\left(\begin{array}{cccc}
0 & 0 & 0 & 0 \\
0 & 0 & 0 & 0 \\
0 & 0 & 0 & 0 \\
0 & 0 & 0 & 2
\end{array}\right),\ \lambda_{00} = 0;\
R_{33} = \frac{1}{\sqrt{2}}\left(\begin{array}{cccc}
1 & 0 & 0 & 0 \\
0 & 0 & 0 & 0 \\
0 & 0 & 0 & 0 \\
0 & 0 & 0 & -1
\end{array}\right),\ \lambda_{33} = -\alpha
\end{equation}
and those in Eq.(14) to Eq.(17), but with the following respective eigenvalues,
$$
\lambda_{01}^{\pm} = \lambda_{02}^{\pm} = \lambda_{03}^{\pm} = -\frac{1}{2}\alpha,
$$
\begin{equation}
\lambda_{11} = \lambda_{12}^{\pm} = \lambda_{13}^{\pm} = \lambda_{22} = \lambda_{23}^{\pm} = 0.
\end{equation}
The left eigenoperators are determined as above, and Eq.(7) becomes
$$
\pi \longrightarrow \rho = \sum_i{\rm tr}(L_i\pi)\exp(\lambda_it)R_i = \sum^1_{j = 0}E^c_{jj}\pi E^{c\dagger}_{jj},
$$
where
\begin{equation}
E^c_{00} = \left(\begin{array}{cccc}
\cos\chi & 0 & 0 & 0 \\
0 & 1 & 0 & 0 \\
0 & 0 & 1 & 0 \\
0 & 0 & 0 & 1
\end{array}\right),\
E^c_{11} = \left(\begin{array}{cccc}
0 & 0 & 0 & 0 \\
0 & 0 & 0 & 0 \\
0 & 0 & 0 & 0 \\
\sin\chi & 0 & 0 & 0
\end{array}\right),
\end{equation}
with
\begin{equation}
\cos\chi \equiv \exp\left(-\frac{1}{2}\alpha t\right),\
\sin\chi \equiv \sqrt{1 - \exp(-\alpha t)}.
\end{equation}
We note that, in contrast to $D^c_{00}$ in Eq.(4), $E^c_{00}$ cannot be written as a tensor product of two two-by-two matrices.  This gives rise to the typical ``spooky'' action of the channel: $|01\rangle$, $|10\rangle$, and any linear combination of them, and $|11\rangle$ will go through the channel undisturbed, but not $|00\rangle$.

It is interesting to note that Eq.(3) can also be derived by solving the following Lindblad equation,
$$
{\cal L}\pi = -\frac{1}{2}\Gamma[\pi - (I \otimes \sigma_z)\pi(I \otimes \sigma_z)] - \frac{1}{2}\Gamma[\pi - (\sigma_z \otimes I)\pi(\sigma_z \otimes I)].
$$
However, analogous approach for the amplitude damping channel with uncorrelated noise does not work.  This is because the amplitude damping channel is by definition non-unital.  This is not surprising in view of the fact that although all Lindblad superoperators have a Kraus decomposition, the converse is not true in general.

Now, we carry out the same analysis as in Ref.\cite{Mac} by considering
\begin{equation}
\pi \longrightarrow \rho = \Phi(\pi) = (1 - \mu)\sum^1_{i, j = 0}E^u_{ij}\pi E^{u\dagger}_{ij} + \mu\sum^1_{i = 0}E^c_{ii}\pi E^{c\dagger}_{ii}
\end{equation}
and using Eq.(2).  Eq.(1) then becomes
$$
I_2 = -\sum^4_{i = 1}t_i\log_2t_i + \frac{1}{4}\sum^4_{j = 1}u_j\log_2u_j 
+ \frac{1}{4}\sum^4_{k = 1}v_k\log_2v_k + \frac{1}{2}\sum^2_{l = 1}w_l\log_2w_l
$$
with
$$
\Theta(\mu,\ \chi) \equiv \frac{1}{2}\left[
(3 + \mu) + (1 - \mu)\left(\cos 4\chi - 32\mu\cos^2\chi\sin^4\frac{\chi}{2}
\right)\right],
$$
$$
t_{1,2} = \frac{1}{4}(1 \pm \sin^2\chi)[(1 \pm \sin^2\chi) \mp \mu\sin^2\chi],
$$
$$
t_{3,4} = \frac{1}{4}[1 - (1 - \mu)\sin^4\chi],
$$
$$
u_{1, 2} = (1 - \mu)\cos^2\theta\cos^2\chi\sin^2\chi,
$$
$$
u_{3, 4} = \frac{1}{2} - (1 - \mu)\cos^2\theta\cos^2\chi\sin^2\chi
\pm \frac{1}{2}\sqrt{\cos^4\theta\cos^22\chi + \sin^4\theta 
+ \cos^2\theta\sin^2\theta\Theta(\mu,\ \chi)},
$$
$$
v_{1, 2} = (1 - \mu)\sin^2\theta\cos^2\chi\sin^2\chi,
$$
$$
v_{3, 4} = \frac{1}{2} - (1 - \mu)\sin^2\theta\cos^2\chi\sin^2\chi
\pm \frac{1}{2}\sqrt{\sin^4\theta\cos^22\chi + \cos^4\theta 
+ \cos^2\theta\sin^2\theta\Theta(\mu,\ \chi)},
$$
\begin{equation}
w_1 = \mu + (1 - \mu)\cos^2\chi,\ w_2 = (1 - \mu)\sin^2\chi.
\end{equation}
Numerical calculation of $I_2$ for $\theta = \frac{\pi}{4}$ (i.e., the maximally entangled Bell states), and $\theta = 0$ (i.e., the completely unentangled product states), with $0 \leq \chi \leq \frac{\pi}{2}$ and $0 \leq \mu \leq 1$ allows us to compare the information carrying capacity of both forms of input state.  Constructing graphs similar to that in Ref.\cite{Mac} shows for each $\chi$ there is a threshold $\mu_t$ such that for $\mu > \mu_t$, the performance of the Bell states for classical information transmission is better than that of the product states.  While, for $\mu < \mu_t$, better performance is achieved by using the product states instead.  For instance, when $\chi = \frac{\pi}{5}$, we have $\mu_t \in (0.5,\ 0.6)$.  Furthermore, for the product states
\begin{equation}
I_2(\theta = 0, \mu = 1, \chi) \geq I_2(\theta = 0,\ \mu = 0, \chi).
\end{equation}
This is a direct consequence of the following inequality:
$$
a\log a + x\log x \leq (a + x)\log(a + x),\ \forall a,\ x \geq 0.
$$
What this shows is that in the case of quantum amplitude damping channel with perfect memory, it is possible to obtain enhanced information carrying performance even if we are using product input states.  This is the same conclusion reached in Ref.\cite{Mac} for the case of a depolarizing channel.

In conclusion, we have extended the problem of time-correlated noise (or ``channles with memory'') as considered in Ref.\cite{Mac} to the case of the amplitude damping channel.  In the case of sending two qubits by successive uses of an amplitude damping channel with partial memory, we establish numerically that by using maximally entangled states rather than product states as information carriers, we can enhance the transmission of classical information over the quantum channel.

The authors thank Yuri Suhov and Daniel Oi for useful discussions.  This publication is an output from project activity funded by The Cambridge MIT Institute Limited (``CMI'').  CMI is funded in part by the United Kingdom Government.  The activity was carried out for CMI by the University of Cambridge and Massachusetts Institute of Technology.  CMI can accept no responsibility for any information provided or views expressed.

\end{document}